\documentclass[12pt]{article}
\usepackage{graphicx}
\usepackage{amssymb,amsmath}
\usepackage{epstopdf,epsfig}
\def\half{\frac{1}{2}}

\newcommand{\tr}{\ensuremath{\text{tr}\,}}

\title{A DK Phase Transition in $q$-Deformed Yang-Mills on $S^2$ and Topological Strings}
\author{Daniel Jafferis and Joseph Marsano {\footnote{E-mail: jafferis@string.harvard.edu, marsano@physics.harvard.edu}}\\
Jefferson Physical Laboratory\\Harvard University\\Cambridge, MA 02138}
\begin{document}

\maketitle

\abstract{We demonstate the existence of a large $N$ phase transition with respect to the 't Hooft coupling in q-deformed Yang-Mills theory on $S^2$. The strong coupling phase is characterized by the formation of a clump of eigenvalues in the associated matrix model of the Douglas-Kazakov (DK) type \cite{Douglas:1993ii}. By understanding this in terms of instanton contributions to the q-deformed Yang-Mills theory, we gain some insight into the strong coupling phase as well as probe the phase diagram at nonzero values of the $\theta$ angle. The Ooguri-Strominger-Vafa \cite{Ooguri:2004zv} relation of this theory to topological strings on the local Calabi-Yau $\mathcal{O}(-p) \oplus \mathcal{O}(p-2) \rightarrow \mathbb{P}^1$ via a chiral decompostion at large $N$ \cite{Aganagic:2004js}, motivates us to investigate the phase structure of the trivial chiral block, which corresponds to the topological string partition function, for $p>2$. We find a phase transition at a different value of the coupling than in the full theory, indicating the likely presence of a rich phase structure in the sum over chiral blocks.}

\begin{section}{Introduction}

One of the most exciting developments in the past few years has been the conjecture of Ooguri, Strominger, and
Vafa \cite{Ooguri:2004zv} relating a suitably defined partition function of supersymmetric black holes to computations in topological
strings.  In particular, they argue that the perturbative corrections to the entropy of four dimensional BPS black holes, in a mixed ensemble, of type II theory compactified on a Calabi-Yau $X$ are captured to all orders by the topological string partition function on $X$ via a relation that takes the form

\begin{equation}Z_{BH}\sim |Z_{top}|^2\label{OSV}\end{equation}

Moreover, since the left-hand side of \eqref{OSV} is nonperturbatively well-defined, it can also be viewed as providing a definition of the nonperturbative completeion of $|Z_{top}|^2$.

One of the first examples of this phenomenon is the case studied in \cite{Vafa:2004qa,Aganagic:2004js} of type II "compactified" on the
noncompact Calabi-Yau ${\cal{O}}(-p)\oplus {\cal{O}}(p-2+2g)\rightarrow\Sigma_g$, with $\Sigma_g$ a Riemann
surface of genus $g$. The black holes considered there are formed from $N$ D4 branes wrapping the 4-cycle
$\mathcal{O}(-p) \rightarrow \Sigma_g$, with chemical potentials for D0 and D2 branes turned on. The mixed entropy
of this system is given by the partition function of topologically twisted $\mathcal{N}=4$ $U(N)$ Yang Mills
theory on the 4-cycle, which was shown to reduce to q-deformed $U(N)$ Yang Mills on $\Sigma_g${\footnote{Before this connection was found, $q$-deformed Yang-Mills had previously been introduced, with different motivations, in \cite{Klimcik:1999kg}.}}.  This theory is defined by the action

\begin{equation}S=\frac{1}{g_{YM}^2}\int_{\Sigma_g}\tr\Phi\wedge F + \frac{\theta}{g_{YM}^2}\int_{\Sigma_g}
\tr\Phi\wedge K-\frac{p}{2g_{YM}^2}\int_{\Sigma_g}\tr\Phi^2\wedge K ,\end{equation}

where $K$ is the Kahler form on $\Sigma_g$ normalized so that $\Sigma_g$ has area 1, $\Phi$ is defined to be
periodic with period $2\pi$, $N$ denotes the large, fixed $D4$-brane charge, and the parameters $g_{YM}^2,\theta$
are related to chemical potentials for $D0$-brane and $D2-$brane charges by

\begin{equation}\phi^{D0}=\frac{4\pi^2}{g_{YM}^2}\qquad \phi^{D2}=\frac{2\pi\theta}{g_{YM}^2}. \label{Dchempots}\end{equation}

The authors of \cite{Aganagic:2004js} went on to verify that the partition function of this theory indeed admits a factorization of a form similar to \eqref{OSV}

\begin{equation}Z=\sum_{\ell\in\mathbb{Z}}\sum_{P,P'}Z^{qYM,+}_{P,P'}(t+pg_s\ell)Z^{qYM,-}_{P,P'}(\bar{t}-pg_s\ell)
\label{tsfact}\end{equation}

where

\begin{equation}\begin{split}Z^{qYM,+}_{P,P'}(t)&=q^{(\kappa_P+\kappa_{P'})/2}e^{-\frac{t(|P|+|P'|)}{p-2}}\\
&\qquad\times\sum_R q^{\frac{p-2}{2}\kappa_R}e^{-t|R|}W_{PR}(q)W_{P^{\prime t}R}(q)\\
Z^{qYM,-}_{P,P'}(\bar{t})&=(-1)^{|P|+|P'|}Z^{qYM,+}_{P^t,P^{\prime t}}(\bar{t})\label{chbldef}\end{split}\end{equation}

and

\begin{equation}g_s=g_{YM}^2\qquad t=\frac{p-2}{2}g_{YM}^2N-i\theta\end{equation}

where $P$, $P'$, and $R$ are $SU(\infty)$ Young tableaux that are summed over.  The extra sums that seem to distinguish \eqref{tsfact} from the conjectured form have been argued to be associated with noncompact moduli and are presumably absent for compact Calabi-Yau.  The chiral blocks $Z^{qYM,+}_{P,P'}$ can be identified as perturbative topological string ampltiudes with 2 "ghost" branes inserted \cite{Aganagic:2004js}.

This factorization is analagous to a phenomenon that occurs in pure 2-dimensional $U(N)$ Yang-Mills, whose partition function can also be written at large $N$ as a product of "chiral blocks" \cite{Gross:1992tu} \cite{Gross:1993hu} \cite{Gross:1993yt}  that are coupled together by interactions analagous to the ghost brane insertions in \eqref{tsfact}.  Indeed, because of the apparent similarity between the pure and $q$-deformed theories, it seems reasonable to draw upon the vast extent of knowledge about the former in order to gain insight into the behavior of the latter{\footnote{A nice review of 2D Yang-Mills can be found, for instance in \cite{Cordes:1994fc}}}.

One particularly striking phenomenon of pure 2-dimensional Yang-Mills theory is a third order large $N$ phase transition for $\Sigma_g=S^2$ that was first studied by Douglas and Kazakov \cite{Douglas:1993ii}.  The general nature of this transition is easy to understand from a glance at the exact solution

\begin{equation}Z_{YM,S^2}=\sum_{n_i\ne n_j}\left[\prod_{i<j}(n_i-n_j)^2\right]\exp\left\{-\frac{g_{YM}^2}{2}\sum_in_i^2\right\}\label{intpymex}\end{equation}

In particular, we see that the system is equivalent to a "discretized" Gaussian Hermitian matrix model with the $n_i$ playing the role of "eigenvalues".  Since the effective action for the "eigenvalues" is of order $N^2$, the partition function is well-approximated at large $N$ by a minimal action configuration whose form is determined by a competition between the attractive quadratic potential and the repulsive  Vandermonde term.  This configuration is well-known to be the Wigner semi-circle distribution and accurately captures the physics of \eqref{intpymex} at sufficiently small 't Hooft coupling.  As the 't Hooft coupling is increased, however, the attractive term becomes stronger and the dominant distribution begins to cluster near zero with eigenvalue separation approaching the minimal one imposed by the "discrete" nature of the model.  When these separations indeed become minimal, the system undergoes a transition to a phase in which the dominant configuration contains a fraction of eigenvalues clustered near zero at minimal separation.  Sketches of the weak and strong coupling eigenvalue densities may be found in figure \ref{sadptsketch}.

As mentioned before, this theory admits a factorization of the form $Z_{YM}=|Z_+|^2$ at infinite $N$ and, moreover, the partition function can be reliably computed in the strong coupling phase as a perturbative expansion in $1/N$ \cite{Gross:1992tu} \cite{Gross:1993hu}  \cite{Gross:1993yt}.   In fact, roughly speaking the chiral blocks correspond to summing over configurations of eigenvalues to the right or left of the minimally spaced eigenvalues near zero.  At the phase transition point, though, it is known that the perturbative expansion breaks down \cite{Taylor:1994zm} so that the perturbative chiral blocks cease to capture the physics of the full theory, even when perturbative couplings between them are included.

A natural question to ask is whether the chiral factorization \eqref{OSV} exhibits similar behavior in any known examples.  The goal of the present work is to lay the groundwork for studying this by first addressing the following fundamental questions.  First, does a phase transition analagous to that of Douglas and Kazakov occur in the $q$-deformed theory on $S^2$?  If so, is there a natural interpretation for the physics that drives it? Is there a correspondingly interesting phase structure of the perturbative chiral blocks that allows one to catch a glimpse of this physics?  Can we extend our study of the phase structure to nonzero values of the $\theta$ angle?

We will find that the answer to the first question is affirmative for $p>2$.  Moreover, we will demonstrate that, as in the case of pure Yang-Mills on $S^2$ \cite{Gross:1994mr}, the transition is triggered, from the weak coupling point of view, by instantons{\footnote{The contribution of instanton sectors in pure Yang-Mills was first written in \cite{Minahan:1993tp}}}.  Using this observation, we will study the theory at nonzero $\theta$ angle and begin to uncover a potentially interesting phase structure there as well.  To our knowledge, this has not yet been done even for pure Yang-Mills so the analysis we present for this case is also new.  We then turn our attention to the trivial perturbative chiral block and find a phase transition at a value of the coupling which differs from the critical point of the full theory.  In particular, for the case $p=3$, which we analyze in greatest detail, the perturbative chiral block seems to pass through the transition point as the coupling is decreased without anything special occurring.

The skeptical reader may wonder whether a detailed analysis is required to demonstrate the existence of a DK type
phase transition in $q$-deformed Yang-Mills as it may seem that any sensible "deformation" of pure Yang-Mills will
retain it.  However, we point out that the $q$-deformation is not an easy one to "turn off".  In fact, from the
exact expression for the partition function of the $q$-deformed theory

\begin{multline} Z_{BH}=Z_{qYM,g}= \\ \sum_{n_i\ne n_j}\prod_{i<j}\left(e^{-g_{YM}^2(n_i-n_j)/2}-e^{-g_{YM}^2(n_j-n_i)/2}\right)^{\chi(\Sigma)} \exp\left[-\frac{g_{YM}^2 p}{2}\sum n_i^2-i\theta \sum n_i\right]\label{intexsol}
\end{multline}

we see that, in order to make a connection with the 't Hooft limit of pure Yang-Mills it is necessary to take the limit $g_{YM}\rightarrow 0$, $N\rightarrow\infty$, and $p\rightarrow\infty$, while holding $g_{YM}^2Np$, which becomes identified with the 't Hooft coupling of the pure theory, fixed.  Because we are interested in finite values of $p$, we need to move quite far from this limit and whether the transition extends to this regime is indeed a nontrivial question.  However, a glance at \eqref{intexsol} gives hope for optimism as again the effective action for the $n_i$ also exhibits competition between an attractive quadratic potential and repulsive term.  We will later see that this optimism is indeed well-founded.

The outline of this paper is as follows.  In section 2, we review the analysis of the Douglas-Kazakov phase transition in pure 2-dimensional Yang-Mills theory with $\theta=0$ that was heuristically described above and initiate a program for studying the phase structure at nonzero $\theta$.  The primary purpose of this section is to establish the methods that will be used to analyze the $q$-deformed theory in a simpler and well-understood context though, as mentioned before, to the best of our knowledge the extension to nonzero $\theta$ is novel.  In section 3, we study the $q$-deformed theory and demonstrate that it indeed undergoes a phase transition for $p>2$ at $\theta=0$ and make steps toward understanding the phase structure at nonzero $\theta$.  In section 4, we turn our attention to the chiral factorization of the $q$-deformed theory as in \eqref{OSV} and study the phase structure of the trivial chiral block, namely the topological string partition function on ${\cal{O}}(-p)\oplus {\cal{O}}(p-2)\rightarrow\mathbb{P}^1$.  We conjecture that, for $p>2$, this quantity itself undergoes a phase transition but at a coupling smaller than the critical coupling of the full $q$-deformed theory.  Finally, in section 5, we make some concluding remarks.

While this work was in progress, we learned that this subject was also under investigation by the authors of \cite{marcos}, whose work overlaps with ours.

\begin{figure}
\begin{center}
\epsfig{file=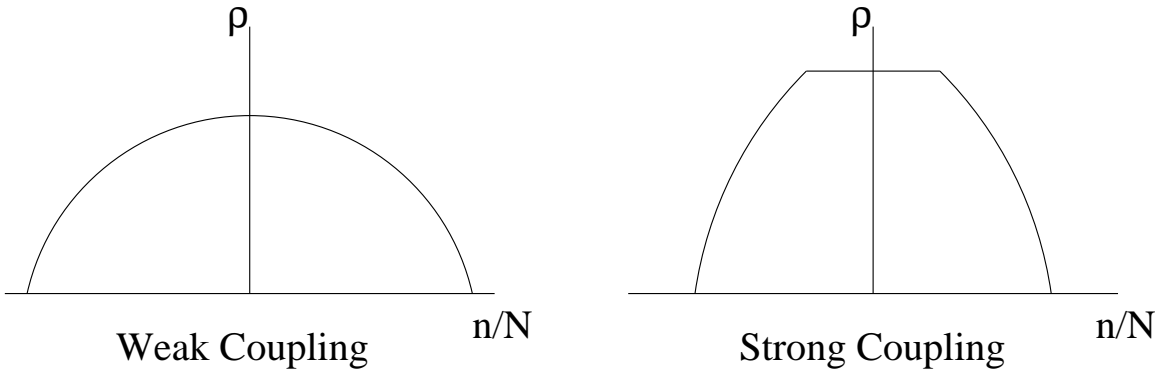,height=5cm}
\caption{Sketch of the dominant distribution of $n_i$ in the weak and strong coupling phases of 2-dimensional Yang-Mills on $S^2$}\label{sadptsketch}
\end{center}
\end{figure}

\end{section}

\begin{section}{The Douglas-Kazakov Phase Transition in Pure 2-D Yang-Mills theory}

In this section, we review several aspects concerning two-dimensional $U(N)$ Yang-Mills theory and the Douglas-Kazakov phase transition with an eye toward an analysis of the $q$-deformed theory in the next section.  In addition to reviewing results for $\theta=0$, we perform a preliminary analysis of the phase structure for nonzero $\theta$ using an expression for the partition function as a sum over instanton contributions.

\begin{subsection}{Review of the Exact Solution}

We begin by reviewing the exact solution of 2-dimensional Yang-Mills.  The purpose for this review is to obtain an expression for the partition function as a sum over instanton sectors that will be useful in our later analysis.  There are many equivalent ways of formulating the theory.  For us, it will be convenient to start from the action

\begin{equation}S=\frac{1}{g_{YM}^2}\int_{S^2}\\tr \Phi\wedge F + \frac{\theta}{g_{YM}^2}\int_{S^2}\tr \Phi\wedge K-\frac{1}{2g_{YM}^2}\int_{S^2}\tr \Phi^2\wedge K\label{ymact}\end{equation}

where $\Phi$ is a noncompact variable that can be integrated out to obtain the standard action of pure 2-dimensional Yang-Mills theory.  To obtain the exact expression \eqref{intpymex}, we first use gauge freedom to diagonalize the matrix $\Phi$, which introduces a Fadeev-Popov determinant over a complex scalar's worth of modes

\begin{equation}\Delta_{FP}=\det([\phi,\ast])\label{FP}\end{equation}

and reduces the action to

\begin{equation}S_{\text{Diag }\Phi}=\frac{1}{g_{YM}^2}\int_{\Sigma}\phi_{\alpha}\left[dA_{\alpha\alpha}-iA_{\alpha\beta}\wedge A_{\beta\alpha}\right]+\frac{\theta}{g_{YM}^2}\int_{\Sigma}\phi_{\alpha}-\frac{1}{2g_{YM}^2}\int_{\Sigma}\phi_{\alpha}^2\label{phidiagact}\end{equation}

Integrating out the off-diagonal components of $A$ in \eqref{phidiagact} yields an additional determinant over a Hermitian 1-form's worth of modes

\begin{equation}\det\,^{,-1/2}_{1F}([\phi,\ast])\label{offdiag}\end{equation}

which nearly cancels the determinant \eqref{FP}.  Combining \eqref{FP} and \eqref{offdiag}, we are left with only the zero mode contributions

\begin{equation}\frac{\Delta(\phi)^{2b_0}}{\Delta(\phi)^{b_1}}=\Delta(\phi)^{2}\label{ec}\end{equation}

where $\Delta(\phi)$ is the usual Vandermonde determinant

\begin{equation}\Delta(\phi)=\prod_{1\le i<j\le N}(\phi_i-\phi_j)\label{vander}\end{equation}

As a result, we obtain an Abelian theory in which each $F_{\alpha}$ is a separate $U(1)$ field strength \cite{Aganagic:2004js}

\begin{equation}Z=\int\,\Delta(\phi)^{2}\,\exp\left\{\int_{S^2}\sum_i\left[\frac{1}{2g_{YM}^2}\phi_i^2-\frac{\theta}{g_{YM}^2}\phi_i-\frac{1}{g_{YM}^2}F_i\phi_i\right]\right\}\label{abelact}\end{equation}

To proceed beyond \eqref{abelact}, let us focus on integration over the gauge field.  The field strength $F_i$ can be locally written as $dA_i$ but this cannot be done globally unless $F_i$ has trivial first Chern class.  We thus organize the gauge part of the integral into a sum over Chern classes and integrations over connections of trivial gauge bundles.  In particular, we write $F_i=2\pi r_i K + F'_i$ where $r_i\in\mathbb{Z}$ and $F'_i$ can be written as $dA_i$ for some $A_i$.  In this manner, the third term of \eqref{abelact} becomes

\begin{equation}-\frac{1}{g_{YM}^2}\int_{\Sigma}\left(2\pi r_iK\phi_{\alpha}+\phi_idA_i\right)\label{Apartt}\end{equation}

Integrating by parts and performing the $A_i$ integral yields a $\delta$ function that restricts the $\phi_i$ to constant modes on $S^2$.  As a result, we obtain the following expression for the pure Yang-Mills partition function organized as a sum over sectors with nontrivial field strength

\begin{equation}Z=\sum_{\vec{r}} Z_{\vec{r}}\label{simpac}\end{equation}

where

\begin{equation}Z_{\vec{r}}=\int\,\prod_{i=1}^N\,d\phi_i\,\Delta(\phi_i)^{2}\exp\left\{-g_{YM}^2\phi_i^2-i\phi_i\left(\theta+2\pi r_i\right)\right\}\label{simpact}\end{equation}

where we have Wick rotated (for convergence) and rescaled (for convenience) the $\phi_i$.
The sum over $r_i$ can now be done to yield a $\delta$-function that sets $\phi_i=n_i$ for integer $n_i$, leaving us with the known result for the exact 2D Yang-Mills partition function

\begin{equation}Z=\sum_{n_i\ne n_j}\Delta(n_i)^{\chi(\Sigma)}\exp\left\{-\frac{g_{YM}^2p}{2}\sum_in_i^2-i\theta\sum_in_i\right\}\label{2dymexact}\end{equation}

As noted in the introduction, this has precisely the form of a "discretized" Hermitian matrix model, with the $\phi_i$ playing the role of the eigenvalues.  In particular, we note that the only effect of summing over $r_i$ in \eqref{simpact} is to discretize the eigenvalues of this matrix model.  Because it is precisely this discreteness that will eventually give rise to the phase transition, it is natural to expect that the transition can also be thought of as being triggered by instantons and the critical point determined by studying when their contributions become nonnegligible.  We will come back to this point later.

Finally we note that the $\theta$ angle has a natural interpretation as a chemical potential for total instanton number.  This is easily seen by shifting $\phi_i$ by $i\theta/g_{YM}^2$ in the contribution from a single instanton sector, \eqref{simpact}, to obtain

\begin{equation}Z_{\vec{r}}=\int\,\prod_{i=1}^N\,d\phi_i\,\Delta(\phi_i)^2\exp\left\{-\frac{g_{YM}^2}{2}\sum_j\phi_j^2+2\pi i \sum_jr_j\phi_j +\frac{2\pi\theta}{g_{YM}^2}\sum_jr_j-\frac{N\theta^2}{2g_{YM}^2}\right\}\label{thchempot}\end{equation}

\end{subsection}

\begin{subsection}{The Phase Transition at $\theta=0$}

In this section, we review the derivation of the transition point in the theory with $\theta=0$ using an analysis that we will eventually generalize to study the $q$-deformed theory.  We can write the exact result \eqref{2dymexact} as

\begin{equation}Z=\sum_{n_i\ne n_j}\exp\left\{-N^2S_{eff}(n_i)\right\}\end{equation}

where

\begin{equation}S_{eff}(n_i)=-\frac{1}{2N^2}\sum_{i\ne j}\ln(n_i-n_j)^2+\frac{\lambda }{2N}\sum_i\left(\frac{n_i}{N}\right)^2\end{equation}

At large $N$, the sums can be computed in the saddle point approximation.  Extremizing $S_{eff}(n_i)$ yields

\begin{equation}\frac{1}{N}\sum_{j;j\ne i}\frac{N}{\phi_i-\phi_j}-\frac{\lambda }{2N}\phi_i=0\label{pymspteqn}
\end{equation}

Following standard methods, we define the density function $\rho$

\begin{equation}\rho(u)=\frac{1}{N}\sum_i\delta\left(u-\frac{\phi_i}{N}\right)\label{pymrhodef}\end{equation}

and approximate $\rho(u)$ by a continuous function at large $N$ for $\phi_i$ distributed in a region centered on zero.  In terms of $\rho$, \eqref{pymspteqn} becomes

\begin{equation}P\int\,dx\,\frac{\rho(x)}{x-u}=-\frac{\lambda}{2}u\label{pymrhoseqn}
\end{equation}

while the fixed number of eigenvalues leads to the normalization condition

\begin{equation}\int\,du\,\rho(u)=1\label{pymrhoncon}\end{equation}

To solve the system \eqref{pymrhoseqn},\eqref{pymrhoncon}, it is sufficient to determine the resolvent

\begin{equation}v(u)=\int_a^b\,dx\,\frac{\rho(x)}{x-u}\label{pymresdef}\end{equation}

as $\rho(u)$ can be computed from $v(u)$ by

\begin{equation}\rho(u)=-\lim_{\epsilon\rightarrow 0}\frac{1}{2\pi i}\left(v(u+i\epsilon)-v(u-i\epsilon)\right)\label{pymrhores}\end{equation}

The resolvent for the system \eqref{pymrhoseqn},\eqref{pymrhoncon} is of course well-known (see for instance \cite{Marino:2004eq})

\begin{equation}v(u)=-\frac{\lambda}{2}\left(u-\sqrt{u^2-\frac{4}{\lambda}}\right)\label{pymressolv}\end{equation}

and leads to the familiar Wigner semi-circle distribution

\begin{equation}\rho(u)=\frac{\lambda}{2\pi}\sqrt{\frac{4}{\lambda}-u^2}\label{pymrhosolv}\end{equation}

As pointed out by Douglas and Kazakov \cite{Douglas:1993ii}, the discreteness of the original sum imposes an additional constraint on the density function, $\rho$, namely

\begin{equation}\rho(u)\le 1\label{pymconstr}\end{equation}

which is satisfied for $\lambda$ such that

\begin{equation}\lambda \le\pi^2\label{pymtrans}\end{equation}

For $\lambda>\pi^2$, \eqref{pymrhosolv} ceases to be an acceptable saddle point of the discrete model.  Rather, the appropriate saddle point in this regime is one which saturates the bound \eqref{pymconstr} over a fixed interval.  Douglas and Kazakov \cite{Douglas:1993ii} performed this analysis and found an expression for the strong coupling saddle in terms of elliptic integrals.  Since the weak and strong coupling saddles agree at $\lambda=\pi^2$, it is clear that the order of the phase transition which occurs at $\lambda=\pi^2$ must be at least second order.  With the saddles in hand, it is also straightforward to compute $F'(\lambda)$, the derivative of the free energy with respect to $\lambda$, as it is simply proportional to the expectation value of $n_i^2$.  On the saddle point, this becomes

\begin{equation}F'(\lambda)\sim N^2\int\,\rho(u)u^2\label{pymfe}\end{equation}

From this, we immediately see that continuity of the saddle point distribution through the critical value also implies that $F'(\lambda)$ is continuous and that, in fact, the transition is at least of second order.  To go further, it is necessary to obtain the strong coupling saddle and evaluate \eqref{pymfe} in both phases.  Douglas and Kazakov have done precisely this and demonstrated that the transition is actually of third order.

\end{subsection}

\begin{subsection}{Instanton Sectors and $\theta\ne 0$}

We now attempt to study the theory with $\theta\ne 0$.  It is not clear to us how to impose the discreteness constraint on the saddle point distribution function $\rho$ once the action becomes complex so we need to probe the phase structure with nonzero $\theta$ by another means.  The key observation that permits us to proceed is that of \cite{Gross:1994mr}, who demonstrated that the phase transition in the $\theta=0$ theory is triggered by instantons in the sense that it occurs precisely when instanton contributions to \eqref{simpac} are no longer negligible at large $N$.  Such a result is not surprising given that it is the discreteness of the model, which arises from the sum over instanton sectors, that drives the transition.

To study what happens with $\theta\ne 0$, we therefore turn to the instanton expansion \eqref{simpac} and ask at what value of the coupling does the trivial sector fail to dominate.  Of course, we must be careful since taking $\theta\rightarrow -\theta$ is equivalent to taking $r_i\rightarrow -r_i$ and shifting $\theta\rightarrow\theta+2\pi k$ is equivalent to shifting $r_i\rightarrow r_i+k$ for integer $k$ so, while the trivial sector dominates near $\theta=0$, the sector $(-n,-n,\ldots,-n)$ dominates at $\theta=2\pi n$.  We avoid any potential ambiguities by restricting ourselves to $0\le\theta\le \pi$.  There, we expect the trivial sector to dominate for $\lambda$ below a critical point at which the first nontrivial instanton sectors, corresponding to $\vec{r}=(\pm 1,0,\ldots,0)$, become nonnegligible.  It is the curve in the $\lambda/\theta$ plane at which this occurs that we now seek to determine.

 We begin with the trivial sector, which is simply the continuum limit of the full discrete model \eqref{2dymexact}.  From \eqref{thchempot}, the partition function within this sector is given by

\begin{equation}Z_{\vec{r}=(0,0,\ldots,0)}=\int\,\prod_{i=1}^N\,d\phi_i\,\exp\left\{\sum_{i<j}\ln(\phi_i-\phi_j)^2-\frac{g_{YM}^2}{2}\sum_i\phi_i^2-\frac{N\theta^2}{2g_{YM}^2}\right\}\end{equation}

The saddle point equation for the $\phi_i$ integral is dominated by the Wigner semicircle distribution found in the previous section \eqref{pymrhosolv}.

We now turn to the family of instanton sectors with $\vec{r}=(r,0,\ldots,0)$ (we will eventually take $r=\pm 1$):

\begin{equation}Z_{\vec{r}=(r,0,\ldots,0)}
=\int\,\prod_{i=1}^N\,d\phi_i\,\exp\left\{-N^2S_{eff,(r,0,\ldots,0)}\right\}
\label{pymrinst}\end{equation}

where

\begin{equation}S_{eff,(\pm 1,0,\ldots,0)}=-\frac{1}{N^2}\sum_{i<j}\ln\left(\phi_i-\phi_j\right)^2+\frac{\lambda }{2N}\sum_i\left(\frac{\phi_i}{N}\right)^2-\frac{2\pi ir}{N}\left(\frac{\phi_1}{N}-\frac{i\theta}{\lambda }\right)-\frac{\theta^2}{2\lambda}\label{pymrinstseff}\end{equation}

To evaluate \eqref{pymrinst}, we note that the saddle point configuration for $\phi_2,\ldots,\phi_N$ in \eqref{pymrinst} is precisely the same as that for the trivial sector since the only difference between the  effective actions in these sectors lie in ${\cal{O}}(N)$ out of the ${\cal{O}}(N^2)$ terms.  We may thus proceed by evaluating the $\phi_2,\ldots,\phi_N$ integrals using \eqref{pymrhosolv} and performing the $\phi_1$ integral explicitly in the saddle point approximation{\footnote{Since we are only interested in the magnitude of the partition function of each sector, it will suffice to determine the real part of the free energy, which is obtained by simply evaluating the action at the potentially complex saddle point.  In particular, we will not need to worry about the precise form of the relevant stationary phase contour.}}.  The effective action for the $\phi_1$ integral becomes

\begin{equation}S_{eff}(\phi_1)=-\int\,dy\,\rho(y)\ln(\phi_1-y)^2+\frac{\lambda}{2}\phi_1^2- 2\pi ir\left(\phi_1-\frac{i\theta}{\lambda }\right)-\frac{\theta^2}{2\lambda }\label{pymrsphon}\end{equation}

and the saddle point value of $\phi_1$ is determined by

\begin{equation}\lambda\sqrt{\phi_1^2-\frac{4}{\lambda}}- 2\pi ir=0\end{equation}

For $\lambda < \pi^2$, there is one saddle point, which lies along the imaginary axis at

\begin{equation}(\phi_1)_{\lambda \le \pi^2r^2}=\frac{2i}{\lambda}\text{sign}(r)\sqrt{r^2\pi^2-\lambda}\end{equation}

As $\lambda$ increases, this saddle point moves toward the real axis, eventually reaching it at $\lambda=\pi^2r^2$ and splitting in two

\begin{equation}(\phi_1)_{\lambda\ge\pi^2}=\pm\frac{2}{\lambda}\sqrt{\lambda-\pi^2r^2}\end{equation}

We now wish to obtain the real part of the "free energy" in these nontrivial sectors by evaluating the effective action \eqref{pymrsphon} on these saddle points.  Comparing with the corresponding free energy of the trivial sector, we obtain

\begin{equation}\begin{split}N^{-1}\left(F_{\vec{r}=(\pm 1,0,\ldots,0)}-F_{\vec{r}=(0,0,\ldots,0)}\right)&=-\frac{2\pi\theta r}{\lambda}+\frac{2\pi^2r^2}{\lambda}\gamma\left(\frac{\lambda}{\pi^2r^2}\right)\qquad\lambda < r^2\pi^2\\
&=-\frac{2\pi \theta r}{\lambda}\qquad\lambda>\pi^2r^2\label{pymfediffs}
\end{split}\end{equation}

where $\gamma(x)$ is defined as in Gross and Matytsin \cite{Gross:1994mr}

\begin{equation}\gamma(x)=\sqrt{1-x}-\frac{x}{2}\ln\left(\frac{1+\sqrt{1-x}}{1-\sqrt{1-x}}\right)=2\sqrt{1-x}\sum_{s=1}^{\infty}\frac{(1-x)^{2s}}{4s^2-1}\label{gammadef}\end{equation}

For $\theta=0$, this result is in agreement with that of \cite{Gross:1994mr} and demonstrates that the partition function of the one-instanton sector is exponentially damped compared to that of the trivial sector at large $N$ for $\lambda\le\pi^2$.  For $\lambda>\pi^2$, on the other hand, these sectors contribute with equal magnitudes at large $N$ and hence instantons are no longer negligible.

In addition this this, we are now able to see at least part of the phase structure with nonzero $\theta$ as well.  While we have only considered a restricted class of instantons here, it is natural to assume that, as $\lambda$ is increased from zero, the most dominant of the nontrivial sectors will continue to be those of instanton number 1 and hence that they will continue to trigger the transition away from $\theta=0$.  We thus arrive at the following phase transition line in the $\lambda/\theta$ plane, which is plotted in figure \ref{gammaplot}

\begin{equation}\frac{\theta}{\pi}=\gamma\left(\frac{\lambda}{\pi^2}\right)\label{pymgthc}\end{equation}

Using the relation between shifts of $\theta$ and shifts of the $r_i$, we can now extend the phase transition line \eqref{pymgthc} outside of the region $0\le\theta\le\pi$.  Moreover, we can attempt to guess the form of the phase diagram beyond this line.  While we know that there are no further transitions at $\theta=0$, it seems likely to us that this is not the case for $\theta\ne 0$.  The reason for this is that for $\theta\ne 0$, it is no longer true that several different sectors contribute equally beyond the transition as is the case for $\theta=0$.  Rather, because of the $\theta$-dependent shift in the free energy, there will in general be one type of instanton sector which dominates all others for any particular value of $\lambda$ and, as $\lambda$ is increased, the instanton number of the dominant sector also increases.  As a result it seems reasonable to conjecture that there are additional transitions that smooth out at $\theta=0$.

We can now ask when sectors of higher instanton number begin to dominate the partition function.  If we assume that the only relevant instantons for determining the full phase diagram are those of the sort considered here, then it is easy to proceed.  However, we find it quite unlikely that the multi-instanton sectors that we have neglected in the present analysis remain unimportant throughout.  To see this, consider following a trajectory at fixed small $\lambda>0$ and along which $\theta$ is increased from 0.  For $\theta$ less than a critical value $\theta_c$ given by \eqref{pymgthc}, the dominant sector is the trivial one.  Just beyond this value, the dominant sector is that with $\vec{r}=(-1,0,\ldots,0)$.  Proceeding further, we expect to hit additional critical points at which a sectors with larger instanton number begin to dominate.  Eventually, though, we will approach $\theta=2\pi-\theta_c$, beyond which the $(-1,-1,\ldots,-1)$ sector becomes the dominant one.  More generally, for each region in the range $0\le\theta\le\pi$ in which an instanton sector $\vec{r}$ dominates, there is a region in the range $\pi\le\theta\le 2\pi$ in which an instanton sector $\vec{r}'=(1,1,\ldots,1)-\vec{r}$ dominates.  A conjecture consistent with this is that a series of transitions occur along this trajectory in which the dominant sectors take the form $(-1,-1,\ldots,-1,0,0,\ldots,0)$.  We sketch a phase diagram based on this conjecture in figure \ref{2dymconjdiag}.  A deeper analysis which takes into account the multi-instanton sectors not considered here in order to confirm/reject a picture of this sort would be very interesting, but is beyond the scope of the present paper.

\begin{figure}
\begin{center}
\epsfig{file=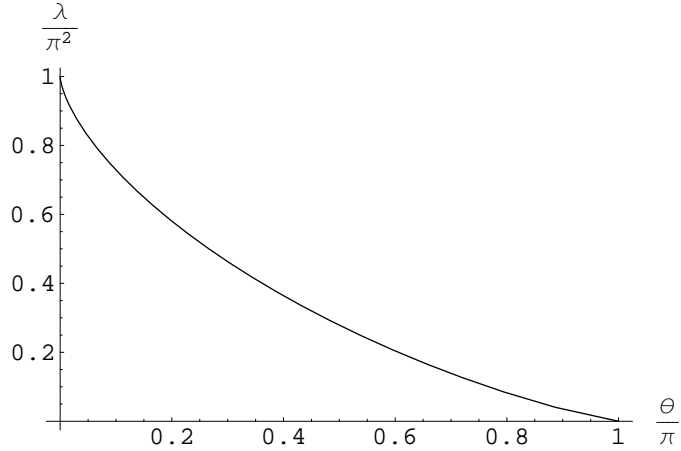,height=6cm}
\caption{Transition line \eqref{pymgthc} in the $\lambda/\theta$ plane}
\label{gammaplot}
\end{center}
\end{figure}

\begin{figure}\begin{center}
\epsfig{file=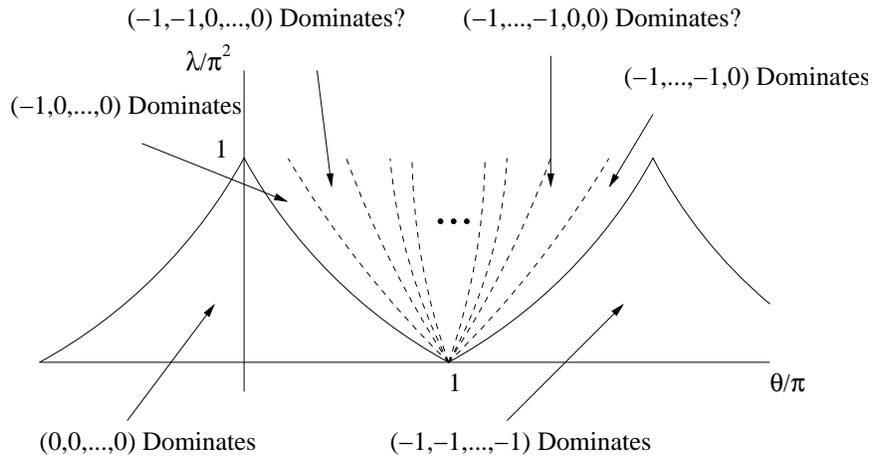,height=6cm}
\caption{Conjecture for the 2D Yang-Mills phase diagram provided we take the curve \eqref{pymgthc} seriously over the entire range  $0\le\theta\le\pi^2$.  The dotted lines represent the first of numerous conjectured transition lines that cluster near $\theta=\pi$.}
\label{2dymconjdiag}
\end{center}\end{figure}

\end{subsection}

\end{section}

\begin{section}{The $q$-deformed Theory}

We now proceed to study the $q$-deformed theory in a manner analagous to the previous section{\footnote{In particular, we will once again use saddle points to study the large $N$ behavior of various instanton sectors.  In pure Yang-Mills, one can obtain more precise results by the technique of  orthogonal polynomials \cite{Gross:1994mr}.  This can in principle be generalized to the $q$-deformed case due to the recent identification of an appropriate set of orthogonal polynomials for such an analaysis  \cite{deHaro:2005rz}}}.  Once again, we begin by reviewing the exact solution while obtaining an expression for the partition function as a sum over instanton sectors.

\begin{subsection}{Review of the Exact Solution}

We begin by recalling the action of the $q$-deformed theory

\begin{equation}S=\frac{1}{g_{YM}^2}\int_{S^2}\tr\Phi\wedge F + \frac{\theta}{g_{YM}^2}\int_{S^2}\,\tr\Phi\wedge K-\frac{p}{2g_{YM}^2}\int_{S^2}\tr\Phi^2\wedge K\label{qymact}\end{equation}

where $\Phi$ is periodic with period $2\pi$.  As demonstrated in \cite{Aganagic:2004js}, the analysis proceeds exactly as in section 2.1 with the periodicity of $\Phi$ giving rise to the replacement

\begin{equation}\Delta(\phi_i)\rightarrow\tilde{\Delta}(\phi_i)=\prod_{1\le i<j\le N}\left[e^{i(\phi_i-\phi_j)/2}-e^{-i(\phi_i-\phi_j)/2}\right]\label{vanrepl}\end{equation}

In particular, \eqref{simpac}-\eqref{simpact} become, after an identical Wick rotation and rescaling

\begin{equation}Z_{qYM}=\sum_{\vec{r}}Z_{qYM,\vec{r}}\end{equation}

with

\begin{multline}Z_{qYM,\vec{r}}=\int\,\prod_{i=1}^N\,d\phi_i\,
\left[\prod_{1\le i<j\le N}\left(e^{-g_{YM}^2(\phi_i-\phi_j)/2}-e^{-g_{YM}^2(\phi_j-\phi_i)/2}\right)^2\right] \\
\times \exp\left\{-\frac{g_{YM}^2p}{2}\phi_i^2-i(\theta+2\pi r_i)\phi_i\right\}\label{qdpartfc}\end{multline}

Performing the sum over the $r_i$ restricts $\phi_i$ to integer values, as in the case of pure Yang-Mills, and yields the known result

\begin{multline}Z_{qYM}=\sum_{n_i\ne n_j}\left[\prod_{1\le i<j\le
N}\left(e^{-g_{YM}^2(n_i-n_j)/2}-e^{-g_{YM}^2(n_j-n_i)/2}\right)\right]^{2} \\ \times
\exp\left\{-\frac{g_{YM}^2p}{2}\sum_in_i^2-i\theta\sum_in_i\right\}\label{qdexact}\end{multline}

As before the partition function resembles a matrix model, this time what appears to be a "discretized" Hermitian matrix model with unitary measure{\footnote{Matrix models of this sort have been studied before in the context of topological strings, first in \cite{Marino:2002fk} and later in \cite{Aganagic:2002wv}.  Their "discretization" has also previously been studied in \cite{deHaro:2005rz}}}.  In addition, the sum in \eqref{qdpartfcn} can again be interpreted as a sum over instanton sectors with the trivial sector corresponding to the continuum limit and capturing all aspects of the full partition function except for the discreteness.  Moreover, by shifting $\phi_i$ by $i\theta/g_{YM}^2p$ in \eqref{qdpartfcn}, we see that the $\theta$ angle continues to carry the interpretation of a chemical potential for total instanton number:

\begin{equation}\begin{split}Z_{qYM,\vec{r}}&=\int\,\prod_{i=1}^N\,d\phi_i\,\left[\prod_{1\le i<j\le N}\left(e^{-g_{YM}^2(\phi_i-\phi_j)/2}-e^{-g_{YM}^2(\phi_j-\phi_i)/2}\right)^2\right]\\
&\qquad\qquad\times\exp\left\{-\frac{g_{YM}^2p}{2}\sum_j\phi_j^2+2\pi i\sum_jr_j\phi_j +\frac{2\pi\theta}{g_{YM}^2p}\sum_jr_j-\frac{N\theta^2}{2g_{YM}^2p}\right\}\label{qdpartfcn}\end{split}\end{equation}

It seems reasonable to expect that the trivial sector will be dominated by a saddle point configuration analagous to the Wigner semicircle distribution for pure Yang-Mills below a critical coupling.  Beyond this point, one might expect the attractive term in the potential to become sufficiently large that this distribution is too highly peaked to be consistent with discreteness, at which point we expect to find a phase transition that, due to its connection with discreteness, can again be thought of as arising from the effects of instantons.

\end{subsection}

\begin{subsection}{The Phase Transition at $\theta=0$}

We now specialize to the case $\theta=0$ for simplicity and determine the saddle point configuration for small values of the coupling.  We will find a distribution analagous to the Wigner semicircle distribution which will, for sufficiently large coupling, violate the constraint arising from discreteness leading to a phase transition analagous to that of Douglas and Kazakov.  To proceed, we write the exact result \eqref{qdexact} as

\begin{equation}Z=\sum_{n_i\ne n_j}\exp\left\{-N^2S_{eff}(n_i)\right\}\end{equation}

where

\begin{equation}S_{eff}(n_i)=-\frac{1}{2N^2}\sum_{i\ne j}\ln\left(2\sinh\left[\frac{\lambda(n_i-n_j)}{2N}\right]\right)^2+\frac{\lambda p}{2N}\sum_i\left(\frac{n_i}{N}\right)^2\label{qdeffact}\end{equation}

We proceed to study this in the saddle point approximation.  The saddle point condition is easily obtained

\begin{equation}\frac{1}{N}\sum_{j;j\ne i}\coth\left[\frac{\lambda(n_i-n_j)}{2N}\right]=\frac{pn_i}{N}\label{qdspteqn}\end{equation}

We again follow standard techniques and introduce a density function $\rho(u)$ according to \eqref{pymrhodef} in terms of which \eqref{qdspteqn} can be written

\begin{equation}P\int\,dy\,\rho(y)\coth\left[\frac{\lambda(x-y)}{2}\right]=px\label{qdrhospt}\end{equation}

and which must satisfy the normalization condition

\begin{equation}\int\,dy\,\rho(y)=1\label{qdrhoncond}\end{equation}

To solve \eqref{qdrhospt} and \eqref{qdrhoncond}, it is convenient to make the change of variables $Y=e^{\lambda y}$.  In this manner, we may rewrite the system \eqref{qdrhospt}, \eqref{qdrhoncond} as

\begin{equation}\begin{split}P\int\,dY\,\frac{\tilde{\rho}(Y)}{Y-X}&=-\frac{p}{2\lambda}\ln\left[Xe^{-\lambda/p}\right]\\
\int\,dY\,\frac{\tilde{\rho}(Y)}{Y}&=1\label{qdtrhosys}\end{split}\end{equation}

where

\begin{equation}\tilde{\rho}(Y)=\tilde{\rho}\left(e^{\lambda y}\right)=\frac{1}{\lambda}\rho(y)\label{qdtrhodef}\end{equation}

As usual, solving this system of integral equations is equivalent to determining the resolvent

\begin{equation}v(X)=\int\,dY\frac{\tilde{\rho}(Y)}{Y-X}\label{qdresdef}\end{equation}

Fortunately, this model has been studied before \cite{Aganagic:2002wv} and the resolvent found to be

\begin{equation}v(X)=\frac{p}{\lambda}\ln\left[\frac{X+1+\sqrt{(X-a_+)(X-a_-)}}{2X}\right]\label{qdressolv}\end{equation}

where

\begin{equation}a_{\pm}=-1+2e^{\lambda/p}\left[1\pm\sqrt{1-e^{-\lambda/p}}\right]\label{qdapmdef}\end{equation}

From this, we find that $\rho(u)$ is nonzero only for $a_-\le e^{\lambda y}\le a_+$, where it takes the value

\begin{equation}\begin{split}\rho(y)
&=\frac{p}{2\pi}\arccos\left[-1+2e^{-\lambda/p}\cosh^2\left(\frac{\lambda u}{2}\right)\right]
\label{qdrhosolv}\end{split}\end{equation}

The appropriate branch of the $\arccos$ function for this solution is that for which $0\le\arccos x<\pi$.
As in the case of pure Yang-Mills, the discreteness of the model imposes an additional constraint

\begin{equation}\rho(u)\le 1\label{qddiscbd}\end{equation}

which is satisfied provided

\begin{equation}\lambda \le\lambda_{crit}= -p\ln\cos^2\frac{\pi}{p}\label{qdlcrit}\end{equation}

For $\lambda>\lambda_{crit}$, \eqref{qdrhosolv} ceases to be an acceptable saddle point of the discrete model and we expect a phase transition to a distribution analagous to the strong coupling saddle of Douglas and Kazakov at this point.  Note that $\lambda_{crit}$ is finite and nonzero only for $p>2$ so we conclude that a phase transition of the Douglas-Kazakov type occurs only for these $p$ in the $q$-deformed model.

To go beyond $\lambda_{crit}$, we must look for a new saddle which saturates the bound \eqref{qddiscbd} over a fixed interval.  As in the pure Yang-Mills case, this strong coupling saddle must agree with that at weak coupling at $\lambda=\lambda_{crit}$.  Moreover, we note that the derivative of the free energy with respect to $\lambda$ is given at large $N$ by

\begin{equation}\begin{split}F'(\lambda)&=N^2\left[-\frac{1}{2}P\int\,dx\,dy\,\rho(x)\rho(y)(x-y)\coth\left(\frac{\lambda(x-y)}{2}\right)+\frac{\lambda p}{2}\int\,dy\,\rho(y)y^2\right]\\
&=-\frac{N^2p}{2}\int\,dy\,\rho(y)y^2\end{split}
\end{equation}

where we have used \eqref{qdrhospt}.  Consequently, we see that, as in pure Yang-Mills theory, continuity of $\rho(y)$ through the transition point guarantees that $F'(\lambda)$ is continuous and hence the transition must be of at least second order.  It seems very likely to us, however, that the transition will continue to be of third order as in the case of pure Yang-Mills.
We include a brief discussion of the necessary tools to study the strong coupling saddle point in Appendix A.  A more complete presentation, as well as further arguments for the third order nature of the transition, can be found in \cite{marcos}.

\end{subsection}

\begin{subsection}{Instanton Sectors and $\theta\ne 0$}

We now turn $\theta$ back on and attempt to study the phase structure.  As in the pure Yang-Mills case, we seek to probe the phase structure  by analyzing the instanton expansion \eqref{qdpartfcn}.  Again, the trivial sector dominates at small $\theta,\lambda$ and we expect the phase transition to occur when the first nontrivial instanton sector, corresponding to $\vec{r}=(\pm 1,0,\ldots,0)$, makes a nonnegligible contribution to the partition function at large $N$.  We now work by analogy and seek to determine the curve in the $\lambda/\theta$ plane that bounds the region in which the trivial sector dominates.

We begin with the trivial sector, which is simply the continuum limit of the full discrete theory \eqref{qdexact}

\begin{multline}Z_{\vec{r}=(0,0,\ldots,0)}= \\ \int\,\prod_{i=1}^N\,d\phi_i\,\exp\left\{\frac{1}{2}\sum_{i\ne j}\ln\left(2\sinh\left[\frac{\lambda(\phi_i-\phi_j)}{2N}\right]\right)^2-\frac{\lambda
p}{2N}\sum_i\left(\frac{\phi_i}{N}\right)^2-\frac{N\theta^2}{2g_{YM}^2p}\right\}\end{multline}

The saddle point distribution for this integral is given by the function $\rho$ found in the previous section \eqref{qdrhosolv}.

We now turn to the family of nontrivial instanton sectors with $\vec{r}=(r,0,\ldots,0)$ (we will later take $r=\pm 1$) whose partition functions take the form

\begin{equation}Z_{\vec{r}=(r,0,\ldots,0)}=\int\,\prod_{i=1}^N\,d\phi_i\,\exp\left\{-N^2S_{eff,(r,0,\ldots,0)}\right\}\label{qdzseff}\end{equation}

with

\begin{equation}S_{eff,(r,0,\ldots,0)}=-\frac{1}{N^2}\sum_{i<j}\ln\sinh^2\left(\frac{\lambda(\phi_i-\phi_j)}{2}\right)+\frac{\lambda p}{2N}\sum_{i=1}^N\phi_i^2-\frac{2\pi ir}{N}\left(\phi_1-\frac{i\theta}{\lambda p}\right)-\frac{\theta^2}{2\lambda p}
\label{qdseff}\end{equation}

The integrals over $\phi_i$ with $i>1$ are dominated by the saddle point \eqref{qdrhosolv} leading to   the following effective action for the $\phi_1$ integral

\begin{equation}S_{eff}(\phi_1)=-\int\,dy\,\rho(y)\ln\sinh^2\left(\frac{\lambda(\phi_1-y)}{2}\right)+\frac{\lambda p}{2}\phi_1^2-2\pi i r\left(\phi_1-\frac{i\theta}{\lambda p}\right)-\frac{N\theta^2}{2\lambda p}\label{qdseffpone}\end{equation}

It is easy to determine the saddle point equation for $\phi_1$ using \eqref{qdrhospt}

\begin{equation}\frac{\lambda}{p}\left[1-2v\left(e^{\lambda\phi_1}\right)\right]=\lambda \phi_1- \frac{2\pi ir}{p}\label{qdspeqn}\end{equation}

where $v$ is the resolvent from \eqref{qdresdef}.  Solving this equation is straightforward, though requires a bit of care.  If we write $r=np+m$ where $n,m$ are integers and $m\in (-p/2,p/2]$, then the solution can be written as

\begin{equation}\begin{split}\phi_1&=\frac{2i}{\lambda}\text{sign}(r)\arctan\left[\sqrt{e^{-\lambda/p}\sec^2\frac{\pi m}{p}-1}\right]+\frac{2\pi i n}{\lambda}\qquad e^{-\lambda/p}\sec^2\frac{\pi m}{p}>1\\
&=\frac{2}{\lambda}\ln\left[e^{\lambda/2p}\cos\frac{\pi m}{p}\left(1\pm\sqrt{1-e^{-\lambda/p}\sec^2\frac{\pi m}{p}}\right)\right]+\frac{2\pi i n}{\lambda}\qquad e^{-\lambda/p}\sec^2\frac{\pi m}{p}<1\end{split}\label{qdsaddls}\end{equation}

where the appropriate branch of $\arctan$ in \eqref{qdsaddls} is $-\pi/2<\arctan x<\pi/2$ and that of $\ln$ is such that it has zero imaginary part for real argument.

As in the case of pure Yang-Mills, the saddle points lies along the imaginary axis for small $\lambda$ and, as $\lambda$ approaches the critical point  $\lambda_{crit}=-p\ln\cos^2\frac{\pi}{p}$, the saddle points corresponding to $r=\pm 1$ hit the origin and splits in two along the real axis as $\lambda$ is increased further.

To study when particular instanton sectors become nonnegligible, we must evaluate the (real part of) the free energy in each sector, namely

\begin{equation}F_{\vec{r}}=\text{Re}\left[-g(\phi_1)+\frac{\lambda p}{2}\phi_1^2-2\pi ir\phi_1-\frac{2\pi r\theta}{\lambda p}\right]\end{equation}

where

\begin{equation}g(\phi_1)=\int\,dy\,\rho(y)\ln\sinh^2\left(\frac{\lambda(x-y)}{2}\right)^2\end{equation}

In Appendix B, we obtain an exact expression for the function $g(\phi)$ and, in addition, simple integral formulae
for the real part of $g(\phi)$ for $\phi$ along the real and imaginary axes. Using the results of Appendix B, we
may write the difference between the real parts of free energies of the $\vec{r}=(r,0,\ldots,0)$ and
$\vec{r}=(0,0,\ldots,0)$ sectors as

\begin{equation}\begin{split} N^{-1}\left(F_{\vec{r}=(r,0,\ldots,0)}-F_{\vec{r}=(0,0,\ldots,0)}\right)&=
\\ \frac{4p}{\lambda}\int_0^{\alpha_0}\,d\alpha\,&\left[\frac{\pi r}{p}-\text{sign}(r)
\arccos\left(e^{-\lambda/2p}|\cos\alpha|\right)\right]+\frac{2\pi r}{\lambda p}\left(2\pi n p-\theta\right)\\
&\qquad\qquad\qquad\qquad\qquad\text{for }e^{\lambda/p}\cos^2\frac{\pi m}{p}<1\\
&=\frac{2\pi r}{\lambda p}\left(2\pi np-\theta\right)\qquad \text{for }e^{\lambda/p}\cos^2\frac{\pi m}{p}>1
\end{split}\label{qddeltaF}\end{equation}

where

\begin{equation}\alpha_0=\text{sign}(r)\arccos\left(e^{\lambda/2p}\left|\cos\frac{\pi m}{p}\right|\right)
\label{qdalphzer}\end{equation}

and the appropriate branch of $\arccos$ is such that $0\le\arccos x\le\pi$.

Let us first look at the case $\theta=0$.  The integrand in \eqref{qddeltaF} is manifestly positive definite for $\alpha_0$ nonzero and $e^{\lambda/p}\cos^2 (\pi m/p)<1$ so the integral is positive definite for $\alpha_0>0$ and 0 for $\alpha_0=0$.  As a result, we see that as the coupling is increased from zero, the trivial sector dominates until the critical point $\lambda=\lambda_{crit}$ \eqref{qdlcrit} is reached, at which point $\alpha_0$ becomes zero for $r=\pm 1$ and the difference in free energies vanishes for the one-instanton sectors $\vec{r}=(\pm 1,0,\ldots,0)$.  Consequently we conclude that, as in the case of pure Yang-Mills, the phase transition at $\lambda=\lambda_{crit}$ \eqref{qdlcrit} is triggered by nontrivial instanton sectors becoming nonnegligible and contributing with equal magnitude to the trivial one at large $N$.

Now, what happens when we turn $\theta$ back on?  It is again reasonable to assume that, as $\lambda$ is increased
from zero, multi-instanton sectors of the type not considered here remain unimportant so that most dominant of the
nontrivial sectors for $\lambda\le\lambda_{crit}$ \eqref{qdlcrit} will continue to be those of instanton number
$\pm$ 1.  This leads us to the following phase transition line in the $\lambda/\theta$ plane, which is plotted for
$p=3$ in figure \ref{qdphaseplot} \footnote{We would like to thank M. Marino for correspondence regarding a
mistake in the plotted curve in version 1, which was the result of a sign error.}

\begin{equation}\frac{\theta}{\pi}=\frac{2p^2}{\pi^2}\int_0^{\alpha_0}\,d\alpha\,\left[\frac{\pi}{p}-\text{sign}(r)\arccos\left(e^{-\lambda/2p}|\cos\alpha|\right)\right]\label{qdthcrit}\end{equation}

The qualitative behavior of this curve is identical to that of the pure Yang-Mills case in that the critical
coupling decreases from $\lambda_{crit}$ \eqref{qdlcrit} at $\theta=0$ to zero at $\theta=\pi$.  Following the
same reasoning applied to pure Yang-Mills, we also expect further transitions beyond the line \eqref{qdthcrit}
beyond which sectors with larger instanton number dominate the full partition function.  The required symmetries
under simultanoues shifts/flips of $\theta$ and $\vec{r}$ then lead us to conjecture a phase diagram of the form
\eqref{2dymconjdiag} in the case of the $q$-deformed theory as well.  It would be interesting to pursue a further
analysis in order to verify or reject this picture.

\begin{figure}\begin{center}
\epsfig{file=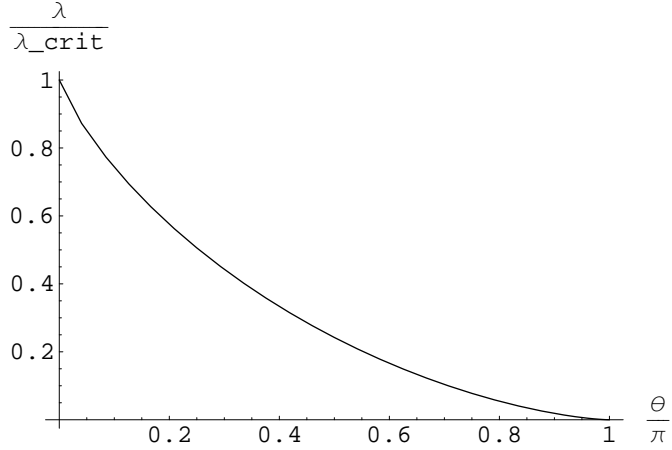,height=6cm}
\caption{Plot of the curve \eqref{qdthcrit} for $r=1,p=3$.}
\label{qdphaseplot}
\end{center}\end{figure}

\end{subsection}

\end{section}

\begin{section}{Chiral decomposition, and phase transition in topological string}

One of the most interesting features of the $q$-deformed Yang-Mills theory we have been examining is its
factorization into chiral blocks at large $N$. This occurs in an analogous manner to the chiral decomposition of
pure Yang Mills, with the new ingredient being its identification by \cite{Aganagic:2004js} with the topological string
partition function. We will show that the chiral partition function has a natural interpretation in the strong coupling phase as a limiting description of the eigenvalues on a single side of the clumped region.

In contrast to the case of Yang-Mills on a torus, where a similar picture was developed \cite{Vafa:2004qa} \cite{Dijkgraaf:2005bp}, here
of course the Fermi surfaces do not decouple, even perturbatively, because of the nontrivial repulsion between the
eigenvalues. We will see that this is encoded in the sum over "ghost" branes found in \cite{Aganagic:2004js}. The presence of the
weak coupling phase adds a new feature to this picture, since at some value of the 't Hooft coupling, the two
Fermi surfaces simply vanish. We will investigate what effect this has on the trivial chiral block.  Throughout this entire section, we consider only the case $\theta=0$ for simplicity.

We now look a bit closer at the trivial chiral block, $Z^{qYM,+}_{..}$, which is to be identified with the
topological string partition function without any ghost brane insertions \cite{Aganagic:2004js}.  Using \eqref{chbldef}, we see that it takes a particularly simple form

\begin{equation}\begin{split}Z_{top}&=\sum_R q^{\frac{p-2}{2}\kappa_R}e^{-t|R|}W_R(q)^2\\
&=\sum_R q^{p\kappa_R/2}e^{-t|R|}W_R(q^{-1})^2\\
\end{split} \label{top} \end{equation}

where the sum is over Young diagrams, $R$ and the objects appearing in the sum are defined as follows:

\begin{equation}\begin{split}\kappa_R&=\sum_iR_i(R_i-2i+1)\\
W_R(q^{-1})&=s_R\left(q^{i-\frac{1}{2}}\right)\\
q&=e^{-g_s}=e^{-g_{YM}^2}
\end{split}
\end{equation}

where $R_i$ denotes the number of boxes in the $i$th row of the diagram $R$ and $s_R(x^i)$ is the $SU(\infty)$ Schur function associated to the diagram $R$.

It is clear that, if we include all Young diagrams, the series \eqref{top} diverges for any $q$ for
$p \geq 3$,
since arbitrarily negative powers of $q$ appear for diagrams with sufficiently large diagrams.
This should not disturb us, however, because the topological
string is merely an asymptotic expansion, and only the perturbation theory about $g_s = 0$ is well defined.  Thus
we truncate the sum over Young diagrams to chiral representations of $U(M)$, for some finite $M$, and take a 't Hooft limit keeping $\lambda = g_s M$ fixed.  In fact, the chiral decomposition \eqref{chbldef} indicates precisely what this cutoff should be as one can easily verify that \eqref{tsfact} holds at finite $N$ only when the Young diagrams $R$ are taken to have at most $N/2$ rows.  To be sure, the model that we study is the chiral block defined in this way.  We then assume that, based on \eqref{tsfact}, this is the proper manner in which to define the perturbative topological string partition function.

\begin{subsection}{Interpretation of the trivial chiral block}

Heuristically, we expect the chiral partition function to describe a system in which only the eigenvalues on one side of the clump in the strong coupling phase are "dynamical" in the sense their locations are summed.  We can isolate such contributions by placing an infinite clump of eigenvalues about the origin which drives the two fermi surfaces apart, effectively decoupling them.  Of course, one must be careful when doing this to avoid introducing divergences in the partition function.  A particularly natural way to proceed is to start with the $q$-deformed Yang-Mills partition function expressed as a sum over $U(N)$ Young diagrams.  This is easily obtained from the exact expression \eqref{intexsol} by the change of variables

\begin{equation}{\cal{R}}_i=n_i-\rho_i\end{equation}

with $\rho$ the Weyl vector of $U(N)$

\begin{equation}\rho_i=\frac{1}{2}\left(N-2i+1\right)\end{equation}

Note that the ${\cal{R}}_i$ parametrize deformations from the maximally clumped rectangular distribution so organizing the partition function as a sum over the ${\cal{R}}_i$ is quite natural for the problem at hand.  The resulting expression is

\begin{equation}Z_{qYM}\sim \sum_{{\cal{R}}}\prod_{1\le i<j\le N}\left[\frac{[{\cal{R}}_i-{\cal{R}}_j+j-i]_q}{[j-i]_q}\right]^2 q^{\frac{p}{2}C_2({\cal{R}})}e^{-i\theta |\cal{R}|}\end{equation}

where the sum is now over $U(N)$ Young diagrams ${\cal{R}}$.  The quadratic Casimir and $|{\cal{R}}|$ are defined as

\begin{equation}\begin{split}C_2({\cal{R}})&=\kappa_{{\cal{R}}}+N|{\cal{R}}|\\
|{\cal{R}}|&=\sum_i{\cal{R}}_i\end{split}\end{equation}

and $[x]_q$ is the $q$-analogue defined by

\begin{equation}[x]_q=q^{x/2}-q^{-x/2}\end{equation}

To add an additional $M$ eigenvalues, we simply pass from $N$ to $M+N$ in the above expression, obtaining a sum over $U(M+N)$ representations that becomes

\begin{equation}\begin{split}Z'&=\sum_{\cal{R}}\left[\prod_{1\le i<j\le M+N}\left(\frac{[{\cal{R}}_i-{\cal{R}}_j+j-i]_q}{[j-i]_q}\right)^2\right]q^{\frac{p}{2}\left([M+N]|{\cal{R}}+\kappa_{{\cal{R}}}\right)}e^{-i\theta |{\cal{R}}|}\\
&=\sum_{\cal{R}}\left[q^{-(N+M-1)|{\cal{R}}|}\prod_{1\le i<j\le M+N}\left(\frac{q^{{\cal{R}}_i-i}-q^{{\cal{R}}_j-j}}{q^{-i}-q^{-j}}\right)^2\right]q^{\frac{p}{2}([M+N]|{\cal{R}}|+\kappa_R}e^{-i\theta |{\cal{R}}|}\\
&=\sum_{\cal{R}}\left[s_{\cal{R}}^{(M+N)}\left(q^{i-\frac{1}{2}}\right)\right]^2q^{\frac{p-2}{2}(M+N)|{\cal{R}}|+\frac{p}{2}\kappa_{\cal{R}}}e^{-i\theta |{\cal{R}}|}
\end{split}\end{equation}

If we now restrict to representations with at most $N/2$ rows, corresponding to treating only the largest $N/2$ eigenvalues as "dynamical", and take the limit $M\rightarrow\infty$, which is well-defined only for $p>2$, we obtain the result

\begin{equation}Z=\sum_R W_R\left(q^{-1}\right)^2e^{-t|R|}q^{p\kappa_R/2}\end{equation}

which is our old friend \eqref{top}.  We thus see the sense in which the chiral block describes the dynamics of eigenvalues to one side of the clump.

\end{subsection}

\begin{subsection}{Phase Structure of the Trivial Chiral Block}

We now perform a preliminary study of the phase structure of the trivial chiral block.  As mentioned before, this
quantity is defined with a cutoff on the sum over representations to those with at most $N/2$ rows so, to study
this model as $g_s\rightarrow 0$, we must work in a 't Hooft limit where $g_{YM}^2N\sim t/2$ is held fixed while
taking $N\sim t/2g_s$ to $\infty$.  It is convenient for this analysis to return to the $n_i$ variables, of which
there are now only $N/2$.  The effective action in these variables becomes

\begin{multline}\frac{S_{eff}}{N^2}= -\frac{1}{2N^2}\sum_{i,j=1}^{N/2}\ln\left(e^{-\lambda n_i/N}-
e^{-\lambda n_j/N}\right)^2+ \\ \frac{2}{\lambda N}\sum_{i=1}^{N/2}Li_2\left(e^{-\lambda
n_i/N}\right)+\frac{p\lambda}{2N}
\sum_{i=1}^{N/2}\left(\frac{n_i}{N}\right)^2+\frac{i\theta}{N}\sum_{i=1}^{N/2}\frac{n_i}{N}
\end{multline}

For simplicity, we now set $\theta=0$.  Proceeding via the usual saddle point method, we find that the configuration which extremizes the action satisfies

\begin{equation}\frac{\lambda}{N}\sum_{j=1}^{N/2}\frac{1}{1-e^{\lambda(n_i-n_j)/N}}+\ln\left(1-e^{-\lambda n_i/N}\right)
+\frac{p\lambda}{2}\left(\frac{n_i}{N}\right)=0\end{equation}

Introducing the eigenvalue distribution $\rho$ as usual, we have

\begin{equation}\int\frac{dy\,\rho(y)}{1-e^{\lambda(x-y)}}+\frac{1}{\lambda}\ln\left(1-e^{-\lambda x}\right)+
\frac{px}{2}=0\end{equation}

where, because the cutoff is at $N/2$ rather than $N$, the normalization condition becomes

\begin{equation}\int\,dy\,\rho(y)=\frac{1}{2}\end{equation}

To solve this, we make the change of variables

\begin{equation}U=e^{\lambda u}\end{equation}

If we then define

\begin{equation}\tilde{\rho}(U)=\frac{2}{\lambda}\rho\left(u=\frac{1}{\lambda}\ln U\right), \end{equation}

then the saddle point equation that we must solve becomes

\begin{equation}P\int\frac{dV\,\tilde{\rho}(V)}{U-V}=\frac{p-2}{\lambda}\ln U+\frac{2}{\lambda}\ln\left(U-1\right)\label{chblsadpt}
\end{equation}

with the constraint

\begin{equation}\int\,\frac{\tilde{\rho}(U)\,dU}{U}=1\label{chblnorm}\end{equation}

Because we expect clumping near zero, we look for a saddle point of the form

\begin{equation}\rho(u)=\begin{array}{ll}1&\quad 0\le u\le a\\ \psi(u)&\quad a\le u\le b\\ 0&\quad u>b\end{array}\end{equation}

We now define the function $\tilde{\psi}$ as

\begin{equation}\tilde{\psi}(U)=\frac{2}{\lambda}\psi\left(u=\frac{1}{\lambda}\ln U\right)\end{equation}

and note that \eqref{chblsadpt} and \eqref{chblnorm} may be written in terms of $\tilde{\psi}$ as

\begin{equation}P \int_A^B\frac{dV\tilde{\psi}(V)}{U-V}=\frac{p-2}{\lambda}\ln U+\frac{2}{\lambda}\ln(S-A)\end{equation}

and

\begin{equation}\int_A^B\frac{dV\tilde{\psi}(V)}{U-V}=-1+\frac{2}{\lambda}\ln A\end{equation}

where

\begin{equation}A=e^{\lambda a}\qquad B=e^{\lambda b}\end{equation}

As usual, we may determine $\tilde{\psi}(U)$ by first computing the resolvent

\begin{equation}f(U)=\int\,dV\,\frac{\tilde{\rho}(V)}{U-V}, \end{equation}

The expression for $f(U)$ is given by the well-known "one-cut" solution (see, for instance, \cite{Marino:2004eq}):

\begin{equation}
f(U)=
\frac{1}{2\pi i}\sqrt{(U-A)(U-B)}\oint_{[A,B]}\,dS\,\frac{\frac{2}{\lambda}\ln(S-A)+ \frac{p-2}{\lambda}\ln
S}{(U-S)\sqrt{(S-A)(S-B)}}
\end{equation}

with integration contour going counterclockwise around the cut $[A,B]$.  To evaluate this, we move the contour away from the cut, and pick up instead the branch cuts of the logarithms
along $[-\infty, 0]$ and $[-\infty, A]$, as well as the pole at $S=U$. The resulting integrals can be easily
evaluated, and we find that

\begin{equation}\begin{split}f(U)&=\frac{2}{\lambda}\ln\left(\frac{U-A}{U}\right)+\frac{p}{\lambda}\ln U+
\frac{p-2}{\lambda}\ln\left(\frac{\sqrt{B(U-A)}+\sqrt{A(U-B)}}{\sqrt{B(U-A)}-\sqrt{A(U-B)}}\right)\\
&\qquad -\frac{p}{\lambda}\ln\left(\frac{\sqrt{(U-A)}+\sqrt{(U-B)}}{\sqrt{(U-A)}-\sqrt{(U-B)}}\right)\end{split}
\end{equation}

Using this, we may determine $\tilde{\psi}(U)$ and, from that, the density $\rho(u)$:

\begin{equation}\rho(u)=\left\{
{\begin{array}{ll}1&0\le e^{\lambda u}<A\\
\frac{1}{\pi}\left[p\arctan\left(\sqrt{\frac{B-e^{\lambda u}}{e^{\lambda u}-A}}\right)-(p-2)\arctan
\left(\sqrt{\frac{A(B-e^{\lambda u})}{B(e^{\lambda u}-A)}}\right)\right]&A\le e^{\lambda u}\le B\\
0&e^{\lambda u}>B\end{array}} \right),\label{tssad}\end{equation}

where we have returned to the original variables.

It remains to fix the position of the cut, $A$ and $B$.  This is done by noting that the normalization constraint
implies $f(0)=-1+\frac{2}{\lambda}\ln A$ while the definition of $f(U)$ implies that $f(U)\sim{\cal{O}}(U^{-1})$
as $U\rightarrow\infty$.  Let's look at the first condition, namely $f(U)=0$.  In the limit $U\rightarrow 0$,
$f(U)$ becomes

\begin{equation}f(U\rightarrow 0)=\frac{2}{\lambda}\ln A - \frac{p}{\lambda}\ln\left(\frac{\sqrt{B}+\sqrt{A}}
{\sqrt{B}-\sqrt{A}}\right)+\frac{p-2}{\lambda}\ln\left(\frac{4AB}{B-A}\right)=-1+\frac{2}{\lambda}\ln
A\end{equation}

On the other hand, in the limit $U\rightarrow\infty$, $f(U)$ becomes

\begin{equation}f(U\rightarrow\infty)=\frac{p}{\lambda}\ln\left(\frac{B-A}{4}\right)+\frac{p-2}{\lambda}
\ln\left(\frac{\sqrt{B}+\sqrt{A}}{\sqrt{B}-\sqrt{A}}\right)\end{equation}

Defining

\begin{equation}X=\frac{\sqrt{B}-\sqrt{A}}{2}\qquad Y=\frac{\sqrt{B}+\sqrt{A}}{2}, \end{equation}

for convenience, we obtain the polynomial equations

\begin{equation}X^{2(p-2)}(Y^2-X^2)=e^{-\lambda/2}\label{feq}\end{equation}

\begin{equation}XY^{p-1}=1\label{seq}\end{equation}

These can be solved easily in principle.  As an example, let us consider the case $p=3$.  From \eqref{seq} we find
$Y=1/\sqrt{X}$.  We now plug into \eqref{feq} and look for real solutions such that $A=(X^{-1/2}-X)^2\ge 1$. There
are four solutions, two of which are real, namely

\begin{equation}X=\frac{\sqrt{\mu}}{2}\left[1 \pm \sqrt{\frac{2}{\mu^{3/2}}-1}\right], \end{equation}

where

\begin{equation}\mu=\frac{4}{\alpha}\left(\frac{2}{3}\right)^{1/3}e^{-\lambda/2}+\frac{\alpha}{2^{1/3}3^{2/3}}
\end{equation}

\begin{equation}\alpha=\left(9+\sqrt{3}e^{-3\lambda/2}\sqrt{27e^{3\lambda}-256e^{3\lambda/2}}\right)^{1/3}
\end{equation}

These solutions are real provided

\begin{equation}\lambda>(2/3)\ln\left(\frac{256}{27}\right)\approx 1.499\end{equation}

but in this range the plus sign leads to $A\le 1$ so it is only the minus solution that will interest us.
Moreover, we find that at
$A=1$ at

\begin{equation}\lambda\approx 1.74174\end{equation}

and hence the density \eqref{tssad} ceases to be an acceptable saddle point of the model below this value of the coupling.  This leads us to conjecture that the $p=3$
chiral block, and presumably the chiral blocks for larger values of $p$ as well, undergoes a phase transition at a critical value of $\lambda$.  Note that, interestingly, this occurs far below the
phase transition of the full $q$-deformed theory, which for $p=3$ corresponds to $\lambda_{crit}=3\ln 4\approx
4.159$.  We plot $\rho(u)$ in this case for various values of $\lambda$ in figure \ref{rhoplot}.
Further study of the phase structure here would be very interesting.  We postpone this to future work.

\begin{figure}\begin{center}
\epsfig{file=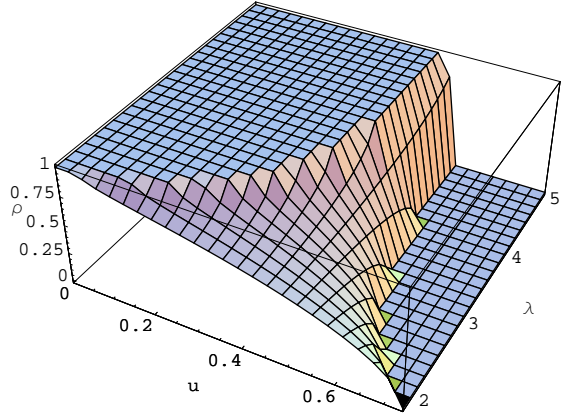,height=6cm}
\caption{The saddle point density $\rho(u)$ as a function of $\lambda$ in the case $p=3$ for $\lambda>\approx 1.74174$.}\label{rhoplot}
\end{center}\end{figure}

\end{subsection}

\end{section}

\begin{section}{Concluding Remarks}

We have found a phase transition in $q$-deformed Yang-Mills theory on $S^2$ in the 't Hooft limit, driven by instantons. Using saddle point techniques we evaluated the free energy in the weak coupling phase. We studied the structure of the contributions of various instanton sectors at strong coupling, and find evidence of a rich phase structure when a $\theta$ angle is turned on. These phase transitions have very interesting consequences for the chiral factorization into a sum over topological string expansions. We discover a phase transition in the trivial chiral block at a different value of the coupling, implying a nontrivial structure in the breakdown of the factorization as one moves into the weak coupling phase.

While in this work we have focused on $p>2$, our analysis also clarifies certain aspects of the OSV relation for $p=1$ and $2$ as well. In those cases, the q-deformed Yang-Mills remains in the weak coupling phase for all $\lambda$, thus the chiral decomposition appears suscipious. In particular, we identify the absence of a clumped region of eigenvalues with the fact that the attractor value of the Kahler class of the $S^2$ is negative for $p=1$. In \cite{Aganagic:2004js}, it is argued that one can perform a flop, which essentially turns on a $\theta$ angle of $-\pi$, in order to make sense of this case.  This is yet another motivation to understand the phase structure at nonzero $\theta$ angle better.    On the other hand, for $p=2$, the attractor size is 0 which is consistent with the fact that only in the limit $\lambda \rightarrow \infty$ do we saturate the discreteness condition.

There are several intriguing directions which deserve further research. It would be very interesting to fully explore the phase structure of the $q$-deformed Yang-Mills for non-zero $\theta$, extending our analysis of the instanton sectors deeper into the strong coupling regime where multi-instanton corrections may become important. This seems particularly promising since we have seen some indictations of a rich phase diagram. The manner in which the chiral decomposition breaks down in the weak coupling phase should also be further elucidated, since we found a phase transition in the topological string partition functions in the trivial sector only after the coupling had moved well inside the regime where we expect the decomposition into chiral blocks to break down. We believe that this phenomenon will be found to be driven by putatively nonperturbative corrections to the factorization becoming important.

\end{section}

\section*{Acknowledgements}

We would like to thank M Marino, L Motl, A Neitzke, K Papadodimas, N Saulina, C Vafa,  and especially S Minwalla for valuable discussions. We are also grateful to M Marino for correspondence prior to publication. The work of DJ was supported in part by NSF grants PHY-0244821 and DMS-0244464. The work of JM is supported was in part by an NSF graduate research fellowship and by DOE grant DE-FG01-91ER40654. DJ would like to thank the Simons Workshop at the Yang Institute of Theoretical Physics at Stonybrook for hospitality.  JM would like to thank the University of Colorado at Boulder and the Perimeter Institute for their hospitality.  Finally, JM would like to thank C Kilic for providing excellent Turkish snacks which were essential in fueling part of this work.

\section*{Appendix A. Two cut solution in the strong coupling phase}

We have seen that for $\lambda > \lambda_{crit}$, the instanton sectors are no longer suppressed, and we have a
transition to a phase in which some of the eigenvalues are clumped together at the maximum density determined by
the constraint \eqref{pymconstr}. Following Douglas and Kazakov \cite{Douglas:1993ii}, we will substitute the ansatz,
\begin{equation} \tilde{\rho} (u)  = 1 \textrm{ for } -a < u < a, \textrm{ and } \rho (u) \textrm{ otherwise},
\label{stcplanatz}\end{equation}
 for the eigenvalue density in the strong coupling regime.

Plugging this into the effective action for q-deformed Yang Mills \eqref{qdeffact}, we find the new effective
action \begin{equation}S_{eff} = \frac{p \lambda}{2} \int du \rho (u) \ u^2 + \frac{p \lambda}{2} \int_{-a}^a dy \
y^2 - \frac{1}{2} \int du \rho (u) \int dv \rho(v) \log \left( \sinh^2 (\frac{\lambda}{2} (u-v)) \right)
\label{stcpeffact}
\end{equation}
$$- \int du \rho (u) \int_{-a}^a dy \log \left( \sinh^2 (\frac{\lambda}{2} (u-y)) \right) -  \frac{1}{2} \int_{-a}^a dx
\int_{-a}^a dy \log \left( \sinh^2 (\frac{\lambda}{2} (x-y)) \right),$$ where we have set $\theta = 0$ at present
for convenience. The saddle point equation is therefore
\begin{equation} W'_{new} (u) = p \lambda u - \log \left( \frac{\sinh^2 \left( \frac{\lambda}{2} (u+a) \right)}{\sinh^2 \left(
\frac{\lambda}{2} (u-a) \right)} \right) = \frac{\lambda}{2} \mathcal{P}\int dy \ \rho(y) \coth \left(
\frac{\lambda}{2} (u - y) \right). \label{stcpsaddle} \end{equation}

This is a system of repulsive eigenvalues in the effective potential, $W_{new}$, which has two wells, one on each
side of the maximally clumped region, that the eigenvalues can fall into. Thus we will look for a symmetric two
cut solution. It will be easier to first change variables to $U = e^{\lambda u}$ and $A = e^{\lambda a}$. This
implies that $$ p \log U -  \log \left( \left( \frac{U^\half A^{\half} - U^{-\half}A^{-\half}}{U^\half A^{-\half}
- U^{-\half}A^\half} \right)^2 \right) = \frac{1}{2} \mathcal{P}\int \frac{dY}{Y} \rho (Y) \frac{U^\half
Y^{-\half} + U^{-\half} Y^\half}{U^{\half} Y^{-\half} - U^{-\half} Y^\half},$$ which can be simplified to
\begin{equation} p \log U - 2 \log \left(\frac{U-A^{-1}}{U-A} \right) + \lambda (\half - 2a) = \mathcal{P} \int
\frac{\rho(Y) dY}{U-Y}. \label{stcpsaddle2} \end{equation}

Defining the resolvent by $ v(U) = \int \frac{ \rho(Y) dY}{U-Y}$, as in equation \eqref{qdresdef}, we are left
with a standard problem of two cut matrix model, see for instance \cite{Marino:2004eq}. The solution is determined
by solving the Riemann-Hilbert problem for the resolvent, which is characterized by falling off as $U^{-1}$ at
infinity, having period $2 \pi i$ about the two cuts, and taking the form $v(U+i \epsilon) - v(U-i \epsilon) = 2
\pi i \rho(U)$ along the cuts. Using well known techniques in complex analysis, one finds that
\begin{multline} v(U) = \\ \oint_{\mathcal{C}} \frac{dz }{2 \pi i} \frac{p \log z - 2 \log \left(\frac{z-A^{-1}}
{z-A} \right) + \lambda (\half - 2a)}{U-z} \sqrt{\frac{(A-U)(A^{-1} - U)(B-U)(B^{-1} - U)}{(A-z)(A^{-1} -
z)(B-z)(B^{-1} - z)}}, \label{rslcontint} \end{multline} where the contour encircles the cuts from $(B^{-1},
A^{-1})$ and $(A, B)$. Moving the contour toward infinity, we pick up instead the branch cut of the logarithm
along $(A^{-1}, A)$, the pole at $z=U$, and the branch cut of the $\log U$ along the negative real axis.

Putting all of this together, we obtain $$ v(U) = p \log U - 2 \log \left(\frac{U-A^{-1}}{U-A} \right) + \lambda
(\half - 2a) + 2 \int_{A^{-1}}^A ds \frac{\sqrt{(A-U)(A^{-1}
 - U)(B-U)(B^{-1} - U)}}{(U-s) \sqrt{(A-s)(A^{-1} - s)(B-s)(B^{-1} - s)}} $$ \begin{equation} - p
\sqrt{(A-U)(A^{-1}
 - U)(B-U)(B^{-1} - U)} \int_{-\infty}^0 ds \frac{1}{(U-s) \sqrt{(A-s)(A^{-1} - s)(B-s)(B^{-1} - s)}}.
 \label{cutsresvl} \end{equation}
This last two integrals can be expressed as elliptic integrals of the third kind. The size of the cut
can be determined by requiring that we reproduce the correct limit as $U \rightarrow \infty$, where we must have $v(U) \sim \mathcal{O}(\frac{1}{U})$, and as $U \rightarrow 0$, where $v(U) = \int \frac{dY}{Y} \rho(Y) = \lambda - 2a$.  Hence we can now compute the free energy in the strong coupling phase.

Define the elliptic integral, $f$, as follows
\begin{equation} \begin{split} f(s) = \int_{A^{-1}}^s \frac{d t}{U-t} \sqrt{\frac{(A-U)(A^{-1}-U)(B-U)(B^{-1}-U)}{(A-t)(A^{-1}-t)(B-t)(B^{-1}-t)}} = \\ \frac{2 i}{B-A} \sqrt{\frac{(B-U)(1-B U)}{(A-U)(1-A U)}} \left\{ (1-AU) F[z|m] + (A^2-1) \Pi [n,z|m] \right\}, \end{split} \end{equation}
where $$z = \sin^{-1} \sqrt{\frac{(B-A)(s-A^{-1})}{(B-A^{-1})(A-s)}} , \ m= \left( \frac{AB-1}{B-A} \right)^2, \textrm{ and } n = \frac{(B-A^{-1})(A-U)}{(B-A)(A^{-1}-U)} . $$

Imposing the limits derived above implies that

\begin{equation} \int_{A^{-1}}^A \frac{d s}{\sqrt{(A-s)(A^{-1}-s)(B-s)(B^{-1}-s)}} = \frac{p}{2} \int_{-\infty}^0 \frac{d s}{\sqrt{(A-s)(A^{-1}-s)(B-s)(B^{-1}-s)}} , \end{equation}
and
\begin{multline} \lambda = p \int_{-\infty}^{-\epsilon} \frac{d s}{s \sqrt{(A-s)(A^{-1}-s)(B-s)(B^{-1}-s)}}
 - p \log \epsilon \\ - 2 \int_{A^{-1}}^A \frac{d s}{s \sqrt{(A-s)(A^{-1}-s)(B-s)(B^{-1}-s)}}, \end{multline}

which can be solved for $A$ and $B$ in terms of $\lambda$. Note that the divergence of the integral up to $-\epsilon$ exactly cancels the $\log \epsilon$.

\section*{Appendix B.  The Function $g(\alpha)$}

In this appendix, we study the function

\begin{equation}g(\alpha)=\int\,dy\,\rho(y)\ln\sinh^2\left(\frac{\lambda(\alpha-y)}{2}\right)\end{equation}

where $\rho(y)$ is the density function \eqref{qdrhosolv} in the strong coupling phase of the $q$-deformed Yang-Mills theory at $\theta=0$.  To start, we note that

\begin{equation}\begin{split}g'(\alpha)&=\lambda\int\,dy\,\rho(y)\coth\left(\frac{\lambda(\alpha-y)}{2}\right)\\
&=\lambda\left[1-2v\left(e^{\lambda\alpha}\right)\right]
\end{split}\end{equation}

where $v$ is the resolvent \eqref{qdresdef} first obtained in \cite{Aganagic:2002wv}.  If we define $A=e^{\lambda\alpha}$, then we can express $g(\alpha)$ as a function of $A$.  Moreover, we have that

\begin{equation}\frac{dg}{dA}(A)=\frac{1}{A}\left(1-2v(A)\right)\end{equation}

which can be integrated in order to obtain $g(A)$:

\begin{equation}\begin{split}g\left(A=e^{\lambda\alpha}\right)&=\ln A+\frac{p}{\lambda}\left[\ln u(A)\right]^2-\frac{2p}{\lambda}\left[\ln u(A)\right]\left[\ln\left(1-e^{-\lambda/p}u(A)\right)\right]\\
&-\frac{2p}{\lambda}Li_2\left(1-u(A)\right)-\frac{2p}{\lambda}Li_2\left(e^{-\lambda/p}u(A)\right)+C\label{gexact}\end{split}\end{equation}

where

\begin{equation}u(A)=\frac{1+A+\sqrt{(A+1)^2-4e^{\lambda/p}A}}{2A}\end{equation}

and $C$ is an $\alpha$-independent constant.  The result \eqref{gexact} is not very useful for learning about free energies.  However, it is useful for noting that

\begin{equation}g\left(\alpha+\frac{2\pi i}{\lambda}\right)=g(\alpha)+\frac{2\pi i}{\lambda}\end{equation}

and hence that shifting $\alpha$ in this manner has no effect on the real part of $g(\alpha)$.  This fact is useful for the analysis described in the main text.

What we really need in order to pursue the analysis within the text is not necessarily $g(\alpha)$ itself, but rather simply $g(\alpha)$ evaluated for $\alpha$ purely real or purely imaginary.  Writing $g$ as $g(z=x+iy)=u(x,y)+iv(x,y)$ we note that

\begin{equation}g'(z)=\partial_z g(z=x+iy)=\partial_xu(x,y)-i\partial_yu(x,y)\end{equation}

and hence that

\begin{equation}u(x,0)=\int\,\text{Re}g'(x,y=0)+D\qquad u(0,y)=-\int\,\text{Im}g'(x=0,y)+D'\end{equation}

where $D,D'$ are constants independent of $x$ and $y$ that depend only on $\lambda$ and $p$.  Computing $u(x,0)$ is particularly easy

\begin{equation}u(x,0)=\frac{\lambda p x^2}{2}\end{equation}

For $u(0,y)$, we have the integral

\begin{equation}u(0,\alpha)=-\lambda\int\,d\alpha\,\left[v\left(e^{i\lambda\alpha}\right)-v\left(e^{-i\lambda\alpha}\right)\right]\end{equation}

which simplifies to

\begin{equation}u(0,\alpha)=-\frac{\lambda p \alpha^2}{2}+2p\int\,d\alpha\,\arctan\left[\sqrt{e^{\lambda/p}\sec^2\frac{\lambda\alpha}{2}}\right]\end{equation}


\begin{thebibliography}{1}

\bibitem{Douglas:1993ii}
  M.~R.~Douglas and V.~A.~Kazakov,
  ``Large N phase transition in continuum QCD in two-dimensions,''
  Phys.\ Lett.\ B {\bf 319}, 219 (1993)
  [arXiv:hep-th/9305047].

\bibitem{Ooguri:2004zv}
  H.~Ooguri, A.~Strominger and C.~Vafa,
  ``Black hole attractors and the topological string,''
  Phys.\ Rev.\ D {\bf 70}, 106007 (2004)
  [arXiv:hep-th/0405146].

\bibitem{Aganagic:2004js}
  M.~Aganagic, H.~Ooguri, N.~Saulina and C.~Vafa,
  ``Black holes, q-deformed 2d Yang-Mills, and non-perturbative topological
  Nucl.\ Phys.\ B {\bf 715}, 304 (2005)
  [arXiv:hep-th/0411280].


\bibitem{Vafa:2004qa}
  C.~Vafa,
  ``Two dimensional Yang-Mills, black holes and topological strings,''
  arXiv:hep-th/0406058.


\bibitem{Klimcik:1999kg}
  C.~Klimcik,
  ``The formulae of Kontsevich and Verlinde from the perspective of the
  Drinfeld double,''
  Commun.\ Math.\ Phys.\  {\bf 217}, 203 (2001)
  [arXiv:hep-th/9911239].


\bibitem{Gross:1992tu}
  D.~J.~Gross,
  ``Two-dimensional QCD as a string theory,''
  Nucl.\ Phys.\ B {\bf 400}, 161 (1993)
  [arXiv:hep-th/9212149].

\bibitem{Gross:1993hu}
  D.~J.~Gross and W.~I.~Taylor,
  ``Two-dimensional QCD is a string theory,''
  Nucl.\ Phys.\ B {\bf 400}, 181 (1993)
  [arXiv:hep-th/9301068].


\bibitem{Gross:1993yt}
  D.~J.~Gross and W.~I.~Taylor,
  ``Twists and Wilson loops in the string theory of two-dimensional QCD,''
  Nucl.\ Phys.\ B {\bf 403}, 395 (1993)
  [arXiv:hep-th/9303046].

\bibitem{Cordes:1994fc}
  S.~Cordes, G.~W.~Moore and S.~Ramgoolam,
  ``Lectures on 2-d Yang-Mills theory, equivariant cohomology and topological
  field theories,''
  Nucl.\ Phys.\ Proc.\ Suppl.\  {\bf 41}, 184 (1995)
  [arXiv:hep-th/9411210].

\bibitem{Taylor:1994zm}
  W.~Taylor,
  ``Counting strings and phase transitions in 2-D QCD,''
  arXiv:hep-th/9404175.

\bibitem{Gross:1994mr}
  D.~J.~Gross and A.~Matytsin,
  ``Instanton induced large N phase transitions in two-dimensional and
  four-dimensional QCD,''
  Nucl.\ Phys.\ B {\bf 429}, 50 (1994)
  [arXiv:hep-th/9404004].

\bibitem{Minahan:1993tp}
  J.~A.~Minahan and A.~P.~Polychronakos,
  ``Classical solutions for two-dimensional QCD on the sphere,''
  Nucl.\ Phys.\ B {\bf 422}, 172 (1994)
  [arXiv:hep-th/9309119].

\bibitem{marcos}
X.~Arsiwalla, R.~Boels, M.~Marino, A.~Sinkovics, "Phase transitions in $q$-deformed 2d Yang-Mills and topological strings," arXiv:hep-th/0509002.

\bibitem{Marino:2004eq}
  M.~Marino,
  ``Les Houches lectures on matrix models and topological strings,''
  arXiv:hep-th/0410165.

\bibitem{deHaro:2005rz}
  S.~de Haro and M.~Tierz,
  ``Discrete and oscillatory matrix models in Chern-Simons theory,''
  arXiv:hep-th/0501123.


\bibitem{Marino:2002fk}
  M.~Marino,
  ``Chern-Simons theory, matrix integrals, and perturbative three-manifold
  Commun.\ Math.\ Phys.\  {\bf 253}, 25 (2004)
  [arXiv:hep-th/0207096].

\bibitem{Aganagic:2002wv}
  M.~Aganagic, A.~Klemm, M.~Marino and C.~Vafa,
  ``Matrix model as a mirror of Chern-Simons theory,''
  JHEP {\bf 0402}, 010 (2004)
  [arXiv:hep-th/0211098].

\bibitem{Dijkgraaf:2005bp}
  R.~Dijkgraaf, R.~Gopakumar, H.~Ooguri and C.~Vafa,
  ``Baby universes in string theory,''
  arXiv:hep-th/0504221.









\end{thebibliography}
 \end{document}